  \documentclass[prl,aps,twocolumn,showpacs]{revtex4}
\usepackage[dvips]{graphicx}
\usepackage{dcolumn}
\newcommand{\be}{\begin{equation}}
\newcommand{\ee}{\end{equation}}
\newcommand{\bea}{\begin{eqnarray}}
\newcommand{\eea}{\end{eqnarray}}
\newcommand{\nn}{ \nonumber}

\newcommand{\ba}{\begin{array}}
\newcommand{\ea}{\end{array}}
\begin{document}
\topmargin=-20mm

\title{On the dissipative effects in the electron transport through
conducting polymer nanofibers }

\author{ Natalya A. Zimbovskaya}\vspace{5mm}

\affiliation
{Department of Physics and Electronics, University of Puerto Rico at
Humacao, Humacao, PR 00791, and Department of Physics and Astronomy, University of Pennsylvania, PA 19104-6323, USA  }

\begin{abstract}
  Here, we study the effects of  stochastic nuclear motions on the electron
transport in doped polymer fibers assuming the conducting state of
the material. We treat conducting polymers as granular metals and
apply the quantum theory of conduction in mesoscopic systems to
describe the electron transport between metalliclike granules.
To analyze the effects of nuclear motions we mimic them by a
phonon bath, and we include electron-phonon interactions in
consideration. Our results show that the phonon bath plays a crucial
part in the intergrain electron transport at moderately low and room
temperatures, suppressing the original intermediate state for the
resonance electron tunneling, and producing new states which support
the electron transport. Also, the temperature dependence of magnitudes of the peaks in the electron transmission corresponding to these new states is analyzed.
\end{abstract}

\date{\today}
\maketitle

\section{i. introduction}

Currently, electron transport in conducting polymer nanofibers is a
subject of intense interest because such nanofibers  are expected to be used in building up nanoelectronic devices. Doped polyacetylen or
polyaniline-polyethylene oxide fibers as well as polypyrrole
nanotubes could behave as conductors, and their conductivity
significantly increases upon doping.  Charge transport mechanisms in
the conducting polymers were intensively studied starting from the
very discovery of these materials \cite{1,2}.

       It is known that chemically doped conducting polymers are very
inhomogeneous. In some regions polymer chains are disorderly
arranged forming an amorphous poorly conducting substance. In other
places the chains are ordered and densely packed \cite{3,4}. The
corresponding regions behave as metalliclike grains embedded in 
the disordered environment. The key point in studies of
the electron transport in conducting polymers is to elucidate the
nature and physical mechanisms of the intergrain transport. Prigodin
and Epstein suggest that the intergrain transport 
mostly occurs due to the electron tunneling  between the grains
through  intermediate resonance states on the polymer chains connecting them. This approach was employed to build up the theory of
electron transport in polyaniline based nanofibers \cite{5}
providing a good agreement with the previous transport experiments
\cite{6}. Assuming the electron tunneling through the intermediate state to be the predominant mechanism for the intergrain transport, we see
a similarity in electron transport mechanisms in conducting polymers
and those in molecules connecting metal leads. In the case of
polymers metalliclike domains take on part of the leads, and
the molecular bridge in between is reduced to a single 
intermediate site.

An important issue that arises in the analysis of the intergrain
electron transport is the influence of nuclear motions in the medium
between the grains. The similar issue was
analyzed while developing the theory of conduction through
macromolecules. It was shown that the environment acts as a source
of incoherence for the tunneling electron \cite{7,8,9,10}. Also, it
can give rise to some extra electron states \cite{11,12}. 
 In the previous works \cite{5,13,14} the effects of stochastic nuclear motions were analyzed by introducing the coupling of the intermidiate state between the grains to a phonon bath including a large amount of harmonic oscillations. It was shown that the direct coupling of the bridge to the dissipative reservoir (phonon bath) brings an inelastic component to the intergrain transport. This erodes the peak in the electron transmission corresponding to the electron tunneling through the bridge. Nevertheless, estimations made in the recent work \cite{15} give grounds to believe that the peak could be still rather well pronounced under typical conditions of experiments on the electrical characterization of conducting polymer nanofibers.

However, the studies of dissipative effects in the intergrain electron transport in conducting polymers may not be restricted with the plain assumption of the direct coupling of the bridge site to the phonon bath. Other scenarios can occur. In particular, analyzing the electron transport in polymers, as well as in 
molecules, one must keep in mind that besides  the bridge sites there always exist other nearby sites. In some cases the presence of such sites may strongly influence the effects of stochastic nuclear motions on the characteristics of electron transport. This may happen when the nearby sites somehow ``screen" the bridge sites from direct interactions with the phonon bath. Here, 
 we elucidate some effects which could appear in the electron transport in conducting polymer fibers in the case of such indirect coupling of the bridge state to the phonon bath.

\section{ii. model and results}

We mimic the effects of
the environment by assuming that the side chain is attached to the
bridge, and this chain is affected by the phonon bath. This model resembles those
used to analyze electron transport through macromolecules
\cite{8,10,17}.  The side chain is introduced to the model to screen
the resonance state making it more stable against the effect of
the phonon bath. As in the previous papers \cite{10,17} we assume
that electrons cannot hop along the side chain, so it may be reduced
to a single site attached to the resonance site (the bridge).

 Within the adopted model the Hamiltonian takes the form:
  \be 
 H = H_{el} + H_{ph} + H_{e-ph} + H_L +H_R
  \ee
 where the Hamiltonian $ H_{el} $ describes the bridge with the attached side site:
  \be 
 H_{el} = \epsilon b^+b +\tilde\epsilon c^+ c - w(b^+ c + h.c.),
  \ee
   $H_{L,R} $ are self-energy terms describing the coupling of the left $(L)$ and right $(R)$ grain to the bridge, and $\epsilon $ and $\tilde\epsilon $ are the on-site energies of the bridge site and the attached side site, respectively. The factor $w$ is the hopping integral between the bridge an the attached site. 

The remaining terms in the expression (1) describe the phonon bath and the electron-phonon interaction, namely:
   \be 
  H_{ph} +H_{e-ph} = \sum_\alpha \Omega_\alpha B_\alpha^+ B_\alpha +
\sum_\alpha \lambda_\alpha c^+ c(B_\alpha + B_\alpha^+)
  \ee
  where $ \Omega_\alpha $ are the frequencies of phonons belonging to the bath and $\lambda_\alpha $ denote the electron-phonon coupling strengths. Performing the unitary transformation $\tilde H= e^S He^{-S} $ with the generator $ S $ \cite{16}:
  \be 
  S =  \sum_\alpha \frac{\lambda_\alpha}{\Omega_\alpha} c^+ c(B_\alpha -B_\alpha^+),
  \ee
  the coupling to the bath is eliminated, and we obtain:
  \be 
  \tilde H = H_{eff} + H_{ph}
  \ee
The effective Hamiltonian for the bridge within the adopted model takes on the form:
  \be 
H_{eff} = \tilde H_{el} + H_L + H_R.
  \ee
 Here,  the Hamiltonian $ \tilde H_{el} $ may be presented as follows:
  \be 
 \tilde H_{el} = H_b + H_c + H_{b-c}
 \ee
 where again $ H_b = \epsilon b^+ b $ corresponds to the bridge site, $ H_c $ is the
Hamiltonian of the side chain, and $ H_{b-c} $ describes the
interaction between them. The terms $ H_c $ and $ H_{b-c} $ are
modified due to the coupling of the side chain to the bath. Keeping
only one site in the side chain we have:
  \be 
 H_c = (\tilde \epsilon - \delta) c^+ c,
   \ee
   \be 
H_{b-c} = - w \chi (b^+ c + h.c.).
  \ee
  The quantities $ \delta $ and $
\chi $ are given by \cite{10}:
 \be 
\delta = \sum_\alpha \frac{\lambda_\alpha^2}{\Omega_\alpha},\qquad
\chi =\exp \bigg[\sum_\alpha \frac{\lambda_\alpha}{\Omega_\alpha}
(B_\alpha - B_\alpha^+)\bigg] .
  \ee

The lowest order in the Fourier transform for the Green's function
associated with the Hamiltonian (6) reads:
  \be 
 G^{-1} (E) = E - \epsilon - \Sigma_L (E) - \Sigma_R (E)- w^2 P(E).
 \ee
 The first four terms in this expression represent the inversed Green's
function for the resonance site coupled to the two grains where $
\Sigma_{L,R} (E) $ are complex self-energy terms. The last term
represents the effect of the environment and has the form \cite{10}:
 \bea 
 P(E)& =&\!\! -i \int_0^\infty dt \exp [it (E - \tilde \epsilon + \delta + i 0^+)]
  \nn \\ & \times &\!\!
\big \{(1 - f) \exp[- F (t)] + f \exp[- F(-t)]\big\}
  \eea
  with $ \exp [-F (t)] $ being a dynamic bath correlation function, and
$ f $ taking on values $1$ and $0 $ when the attached site is occupied and empty, respectively.

Characterizing the phonon bath with a continuous spectral density $ J
(\omega) $ given by \cite{18}:
 \be 
 J (\omega) = J_0 \left(\frac{\omega}{\omega_c}\right )
 \exp \left [-\frac{\omega}{\omega_c} \right],
  \ee
 one may write out the following expressions for the functions $ F (t) $ and $\delta: $ 
  \be 
F (t) = \int_0^\infty \frac{d\omega}{\omega^2} J(\omega)
\left[1 - e^{- i\omega t} + \frac{2[1 - \cos (\omega t)] }{ \exp
(\omega/\theta)-1}\right],
  \ee 
  \be 
\delta = \int_0^\infty \frac{d\omega}{\omega} J(\omega) = J_0
 \ee
 where $ \theta $ is the temperature expressed in the units of energy.

Within the short time scale $(\omega_c t \ll 1) $ the function $ F(t) $ could be presented in the form:
  \be 
 F(t) = \frac{J_0}{\omega_c}\left \{\frac{1}{2}\ln [1 + (\omega_c t)^2]
+ i\arctan (\omega_c t) + K (t) \right \}
   \ee
  where
   \be 
 K(t) = \theta^2 t^2 \zeta \left(2; \frac{\theta}{\omega_c} + 1\right) \equiv
 \theta^2 t^2 \sum_{n=1}^\infty \frac{1}{(n + \theta/\omega_c)^2}.
  \ee
 Here, $ \zeta\ \big(2; \theta/\omega_c + 1\big) $ is the Riemann
$ \zeta $ function \cite{19}.
 The asymptotic expression for $ K(t)$ depends on the relation
between two parameters, namely, the temperature $ \theta $ and the cut-off
frequency  $ \omega_c $ characterizing the thermal
relaxation rate of the phonon bath. 
Assuming $\theta \gg \omega_c $ and  estimating the sum of the series in Eq. (12) and we get:
  \be 
 K(t) \approx \frac{\theta}{\omega_c} (\omega_c t)^2 .
  \ee
In the opposite limit $ \omega_c \gg \theta $ we obtain:
  \be 
 K(t) \approx \frac{\pi^2}{6} (\theta t)^2
  \ee
Also, we may roughly estimate $K(t)$ within the intermediate range. Taking $ \theta \approx \omega_c  $
we arrive at the approximation $K(t) \approx a^2 (\theta t)^2 $ where
$ a^2 $ is a dimensionless constant of the order of unity. 
   Correspondingly, within the short time scale we can omit the first term in the expression (16), and we get:
  \be 
  F(t) \approx \frac{J_0}{\omega_c} \big\{i\omega_c t + K(t)\big\}
  \ee
 where $K(t) $ is given by either Eq. (18) or Eq. (19) depending on the relation between $ \omega_c $ and $ \theta .$

Within the long time scale $ \omega_c t \gg1, $ and provided that temperatures are not very low $(\theta \gg \omega_c),$ we may present the function $ K(t) $ as:
   \be 
  K(t) = 2 \theta t \int_0^\infty \frac{dz}{z^2} (1-\cos z)e^{-z/\omega_c t} \approx \pi\theta t.
  \ee
  Now, the term $K(t) $ is the greatest addend in the expression for $ F(t),$ so the latter could be approximated as: $F(t)\approx \pi\theta t J_0/\omega_c. $ The same approximation holds within the low temperature limit when $\omega_c\gg \theta \gg t^{-1}.$

Using the asymptotic expression (20), we may calculate the contribution to $ P(E) $ coming from the short time scale $(\omega_c t \ll 1).$ It has the form:
  \be 
 P_1(E) =- \frac{i}{2}\sqrt{\frac{\pi}{J_0 \theta}} \exp\left[-
\frac{(E - \tilde\epsilon)^2}{4 J_0\theta} \right] \left \{1+ \Phi
\left[\frac{i(E - \tilde\epsilon)}{2\sqrt{J_0 \theta}} \right]
\right \}
  \ee
  where $ \Phi (z) $ is the probability integral. When both $\omega_c $ and $\theta $ have the same order of magnitude the expression for
$P(E) $ still holds the form (22). At $\theta \ll \omega_c, $ the temperature $ \theta $ in the expression (22) is to be replaced by $ \omega_c. $
 We remark that under the assumption $ \theta \gg \omega_c $ the function $ P_1(E) $ does not depend on the cut-off frequency $\omega_c, $ whereas at $\omega_c \gg
\theta $ it does not depend on temperature.
 The long time $(\omega_c t \gg 1)$ contribution to $ P(E) $ could be similarly estimated as follows:
  \be 
  P_2(E) = \frac{1}{E-\tilde \epsilon + J_0 + i\pi J_0 \theta /\omega_c}.
  \ee
  Comparing these expressions (22) and (23) we see that the ratio of the peak values of $ P_2(E) $ and $ P_1(E) $ is of the order of $(\omega_c^2/J_0\theta)^{1/2}.$ Therefore the term $P_1(E) $ predominates over $P_2(E)$ when the temperatures are moderately high $(\omega_c <\theta)$ and the electron-phonon interaction is not too weak $ J_0/\omega_c \sim1. $ Usually, experiments on the electrical characterization of conducting polymer nanofibers are carried out at $T\sim 100\div300K,$ so in further analysis we assume that $(\omega_c^2/J_0\theta)^{1/2}\ll 1,$ and we omit the term $P_2(E) $ from the following consideration using the approximation $ P(E) \approx P_1(E).$
As shown in the Fig. 1, the imaginary part of $ P(E) $ exhibits a dip around $ E = \tilde\epsilon $ and the width of the latter is determined by the product of the temperature $\theta $ (or $ \omega_c) $ and the constant $ J_0 $ characterizing
the strength of the electron-phonon interaction. When either factor increases, the dip becomes broader and its magnitude reduces.

\begin{figure}[t]
\begin{center}
\includegraphics[width=6.5cm,height=7cm]{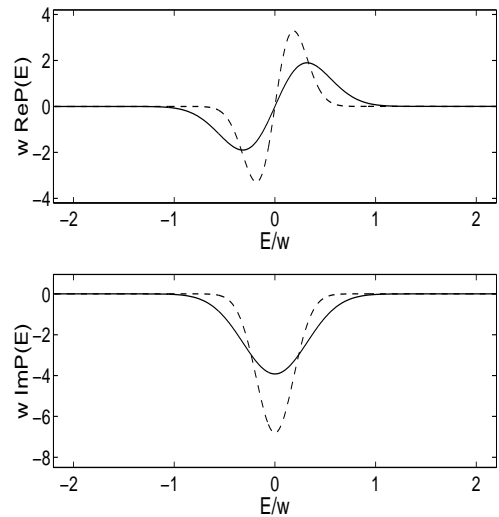}
\caption{Energy dependence of the real (top panel) and imaginary
(bottom panel) parts of $ P(E) .$ The curves are plotted assuming $ J_0 =
20 meV, \ w = 100 meV,\ \tilde \epsilon = 0,\ T= 100 K$ (dash
lines), and $ T = 300 K $ (solid lines) at $ \theta \gg \omega_c. $ } 
\label{rateI}
\end{center}
\end{figure}

The presence of $ P(E) $ gives rise to very significant changes in
the behavior of the Green's function given by the Eq. (11). Using the
flat band approximation for the self-energy corrections $
\Sigma_{L,R} ,$ namely: $ \Sigma_{L,R} = -i\Delta_{L,R} $ where
$\Delta_{L,R} $ are constants of the dimensions of energy, and
neglecting for a moment all imaginary terms in the Eq. (11), we find
that two extra poles of the Green's function emerge due to the
environment. When the phonon bath is detached $(J_0 = 0 )$ the separation between the poles is determined with the hopping integral $ w. $ Due to the electron coupling to the phonons the positions of the poles become shifted.
Assuming $ \epsilon = \tilde \epsilon = 0 $ and $\theta \gg \omega_c $ we get:
  \be 
 E = \pm 2 \sqrt{J_0 \theta |\ln(2J_0 \theta/w^2)|}.
  \ee
  The poles correspond to extra electron states  which
occur due to electrons coupling to the environment. These new states are revealed in the structure of the electron
transmission $ T(E) $ which reads \cite{20}:
  \be 
 T(E) = 4 \Delta_L \Delta_R |G(E)|^2.
 \ee
  The structure of $T(E) $ is shown in the Fig. 2. Two peaks in the
transmission are associated with the environment induced electronic
states. Their positions and heights depend on the temperature  and on the coupling strengths $   J_0, $ and  $ w. $ The important feature in the electron transmission is the absence of the peak associated with the
resonance state between the grains (the bridge site) itself. This
happens due to the strong suppression of the latter by the effects
of the environment. Technically, this peak is damped for it is
located at $ E = 0 $ where the imaginary part of $ P(E) $ reachs its
maximum in magnitude. 
To provide the damping of the original resonance the contribution from the environment (including the side chain attached to the bridge) to the Green's function (11) must exceed the terms $\Sigma_{L,R} $ describing the effect of the grains. This occurs when the inequality
   \be 
  \Delta<w^2/\sqrt{J_0\theta}
  \ee
  is satisfied. When the coupling of the bridge to the attached side site is weak, the influence of the environment slackens and the original peak associated with the bridge at $ E=\tilde \epsilon $ may emerge. At the same time the features originating from the environment induced status become small compared to this peak.

So, the effects of the environment may lead to the
damping of the original resonance state for the electron tunneling
between the metallic islands in the polymer fiber. Instead, two
environment induced states appear to serve as intermediate states
for the electron transport. Similar effects were recently
investigated in the electron transport through DNA macromolecules
\cite{10}, and it was shown that low biased current-voltage
characteristics may be noticeably changed due to the occurence of
the phonon bath induced electron states. Due to some particular
features of conducting polymers, these effects may be revealed in polymer nanofiber current-voltage curves at significantly higher bias voltage, as discussed below.

\begin{figure}[t]
\begin{center}
\includegraphics[width=6.5cm,height=7cm]{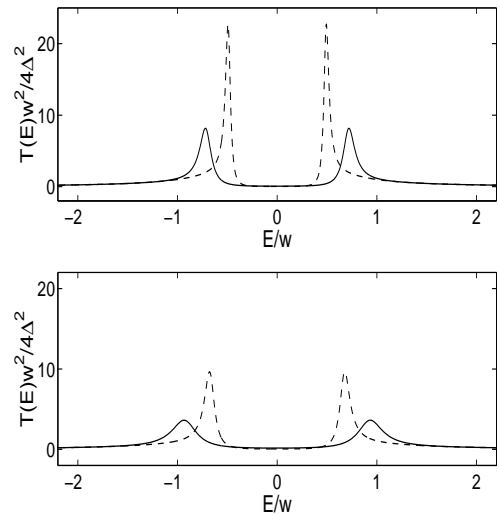}
\caption{ The renormalized electron transmission function vs energy.
The curves are plotted for $ \epsilon = \tilde \epsilon = 0,\ w =
100 meV,\ T = 100 K $ (dash lines) and $ T = 300 K $ (solid lines).
The constant $ J_0 $ equals $ 20 meV $ (top panel) and $ 50 meV $
(bottom panel). } \label{rateI}
\end{center}
\end{figure}

Realistic polymer nanofibers have diameters within the range $
20\div 100$nm, and lengths of the order of a few microns. This is
much greater than the typical size of both metalliclike grains and
intergrain separations which take on values $ \sim 5\div 10$nm
\cite{21}. Therefore, we may treat a nanofiber as a set of parallel
working channels, any single channel being a sequence of grains
connected with the resonance polymeric chains.
 The net current in the fiber is the sum of currents flowing in these
channels, and the voltage $ V$ applied across the whole fiber is
distributed among sequental pairs of grains along a single channel.
So, the voltage $ \Delta V$ applied across two adjacent grains could
be estimated as $ \Delta V \sim V L/L_0 $ where $ L $ is the average
separation between the grains, and $ L_0 $ is the fiber length. The
ratio $ \Delta V/V $ may take on values of the order of $
10^{-1}\div 10^{-2}. $  Experiments on the electrical characterization of the polymer fibers are usually carried out at moderately high temperatures $(T \div 300 K)$, so it seems likely that $ \theta > \omega_c. $
Assuming that $ w \sim 100 meV, $ and $ J_0
\sim 20\div 50 meV $ we estimate the separation between the
environment induced peaks in the electron transmission as $ 120\div
170 meV $ (see Fig. 2). This estimate is close to $ e \Delta V $
when $ V $ takes on values up to $2 \div 3 $ volts. So, the environment
induced peaks in the electron transmission determine the shape
of the current-voltage curves even at reasonably high values of the
bias voltage applied across the fiber.

In calculations of the current we employ the formula \cite{20}:
  \be 
 I = \frac{2en}{h} \int_{-\infty}^\infty dE T(E) [f_1(E) - f_2(E)].
  \ee
  Here, $ n $ is the number of the working channels in the fiber, $ f_{1,2}
(E) $ are Fermi functions taken with the different contact chemical
potentials $ \mu_{1,2} $ for the grains. The chemical potentials
differ due to the applied bias voltage $ \Delta V :$
  \be 
 \mu_1 = E_F + (1 - \eta) e \Delta V; \qquad
 \mu_2 = E_F - \eta e \Delta V.
  \ee
  The parameter $ \eta $ characterizes how the voltage $ \Delta V $ is divided
between the grains; $ E_F $ is the equilibrium Fermi energy of the
system including the pair of grains and the resonance site in
between, and $ T (E) $ is the electron transmission function given
by Eq. (25).

\begin{figure}[t]
\begin{center}
\includegraphics[width=8.9cm,height=4.7cm]{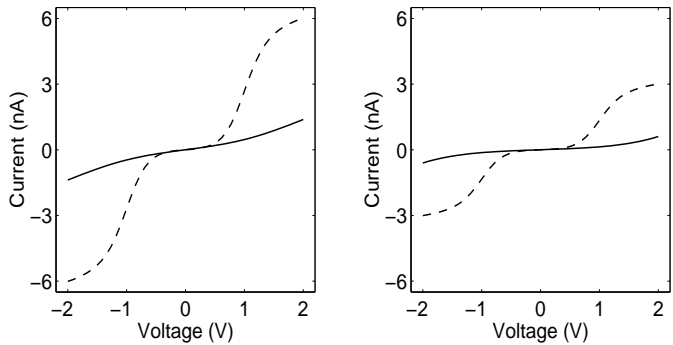}
\caption{
The current-voltage characteristics (nA--V) plotted for $ n =10,\ \ \Delta_L = \Delta_R
= 0.5 meV,\ \ \eta = 1/2,\ \ w = 100 meV,\\ J_0 = 20 meV $ (left panel), and
$ J_0 = 50 meV $ (right panel) at $ T = 100 K $ (dash lines) and $ T = 300K $
(solid lines).}
\label{rateI}
\end{center}
\end{figure}

The resulting current-voltage characteristics are shown in the Fig.
3. The current takes on low values $ (\sim 1 nA) $ because the
coupling of the grains to the intermediate state is weak due to
comparatively large distances between the grains \cite{2}.
Consequently, $ \Delta_{L,R}$ take on values much smaller than those
typical for electron transport through molecules placed between
metal leads. The $ I-V $ curves exhibit a nonlinear shape even at
room temperature despite the fact that the original state for the
resonance tunneling is completely suppressed. This occurs for the
intergrain transport is supported by new phonon induced electron
states. In this work we did not take a task of providing a
quantitative agreement between the theory and experimental results
obtained on conducting polymer fibers. However, in general, our
calculated $ I-V $ curves agree with the reported experiments
\cite{7}.

\begin{figure}[t]
\begin{center}
\includegraphics[width=5.5cm,height=6cm]{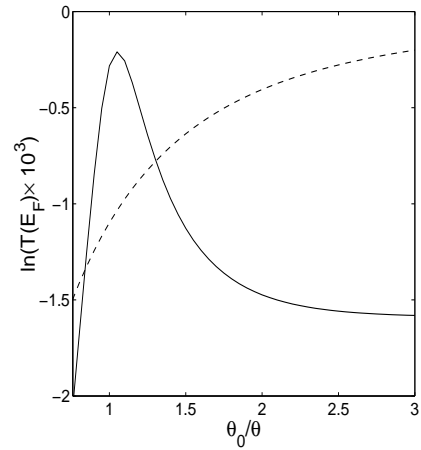}
\caption{Arrhenius plot of the peak value of the electron transmission function for $J_0 =20meV,\ w=100 meV,\ \Delta_L=\Delta_R= 0.5meV;\ \theta_0 = 8.6meV\ (T=100K).$ Solid line is plotted assuming the indirect coupling of the bridge to the phonon bath (Eqs. (11), (25)). Dashed line is plotted assuming that the bridge is directly coupled to the phonons (Eq. (29)).} 
\label{rateI}
\end{center}
\end{figure}

It is worthwhile to discuss the temperature dependence of
the peak value of the electron transmission which follows from the present results. 
Using the expression (11) for the Green's function and the expression (22) for $ P(E) $ we may conclude that at low temperatures the transmission accepts small values, and exhibits rather weak temperature dependence. At higher temperatures $(T\sim 100K) $ the transmission increases fast as the temperature rises and then it reduces as the temperature further increases. The peak in the transmission is associated with the most favorable conditions for the environment induced states to exist when all remaining parameters (such as $J_0 $ and $w)$ are fixed. At high temperatures the peaks associated with the environment induced states are washed out, as usual. 
This is shown in the Fig. 4. Also, basing on the results of the previous works \cite{5,15} we may analyze the temperature dependence of the electron transmission function assuming that the bridge between two adjacent grains is directly coupled to the phonon bath. In this case the transmission peak value may be presented in the form \cite{15}:
   \be 
  T(E) = 1 - \frac{\rho^4}{2(1+ \sqrt{1+\rho^2})^2}
  \ee
  where
   \be 
  \rho^2 = \frac{16J_0}{\omega_c} \frac{\theta^2}{(\Delta_L + \Delta_R)} \zeta \left(2;\frac{\theta}{\omega_c} +1\right) .
   \ee
  The temperature dependence of the transmission given by the Eq. (29) is also shown in the Fig. 4. Both curves are plotted at the same value of the electron-phonon coupling strength $ J_0. $ Comparing them we conclude that at higher temperatures the curves appear to differ. While the temperature rises, we observe a peak in the electron transmission assuming the indirect coupling of the bridge to the phonon bath, and we see the transmission to monotonically decrease when we consider the bridge directly coupled to the bath. Correspondingly, we may expect qualitative diversities to appear in the temperature dependencies of the current, as well. These diversities originate from the difference in the effects of environment on the intergrain electron transport in the cases of direct and indirect coupling of the bridge site to the phonon reservoir. When the bridge is directly coupled to the bath the stochastic motions in the environment only cause the peak in the electron transmission to be eroded, and the higher is the temperature the less distinguishable is the peak. However, when the bridge is screened from the direct coupling with the phonons due to the presence of the nearby sites,
the stochastic nuclear motions in the medium between the grains (especially those in the resonance chain) may take a very different part in the electron transport in conducting polymers at moderately low and room temperatures. Due to their influence the original intermediate state for the resonance tunneling may be completely suppresed but new environments induced states may appear providing the electron transport through the polymer nanofibers.

\section{iii. conclusion}

 Finally, studies of the electron transport in conducting polymers are not completed so far. There are solid grounds to expect significant dissipative effects in the intergrain transport at moderately high temperatures. As shown in the previous studies of the electron transport through molecules, various dissipative effects may occur depending on  characteristic features in the interaction of a propagating electron with the environment. Among these features we may single out the character of the electron coupling to the dissipative reservoir (phonon bath) as a very significant factor. It is likely that in realistic conducting polymers both direct and indirect coupling of the intermediate state (the bridge) to the environment may occur. We need more experimental data concerning transport in these materials to determine which kind of coupling would prevail in a particular material under particular experimental conditions. The model adopted in the present work predicts that the temperature dependence of the electron transmission through the environment induced states crucially differs from that which is typical for the electron transport through the original bridge state directly coupled to the dephasing reservoir. So, the present results give means to experimentaly study distinctive features of the dissipative effects in the electron transport through conducting polymer nanofibers.

\section{Acknowledgments}
 Author thank M. L. Klein and A. T. Johnson Jr. for useful discussions and  G. M. Zimbovsky for help with the manuscript. This work was supported  by NSF Advance program SBE-0123654 and PR Space Grant NGTS/40091.

\end{document}